\documentstyle[epsfig,12pt]{article}
\newcommand{\dd}{{\mathrm d}}
\newcommand{\Li}{{\mathrm Li}_2}
\newcommand{\vecc}[1]{\mbox{\boldmath $#1$}}
\title{Hadronic cross sections in electron-positron
 annihilation with tagged photon}

\author{A.B. Arbuzov$^{1)}$, E.A. Kuraev$^{1)}$, N.P. Merenkov$^{2)}$,\\
and L. Trentadue$^{3)}$}

\date{}

\begin{document}
\maketitle

\begin{center}
{
$^{1)}$ \it Joint Institute for Nuclear Research,
Dubna, 141980, Russia \\
$^{2)}$ \it Institute of Physics and Technology,
Kharkov, 310108, Ukraine \\
$^{3)}$ \it Dipartimento di Fisica, Universit\`a di Parma and
INFN, \\ Gruppo Collegato di Parma, 43100 Parma, Italy
} \\[.2cm]
\end{center}

\begin{abstract}
Events with tagged photons in the process of electron--positron
annihilation into hadrons are considered.
The initial state radiation is suggested
to scan the hadronic cross section with the energy.
QED radiative corrections are taken into account. The results
for the total and exclusive cross sections are given in an
analytic form. Some numerical estimates are presented.
\end{abstract}

\vspace{.6cm}
PACS~ 13.65.+i Hadron production by electron--positron annihilation, \\
12.15.Lk Electroweak radiative corrections

\section{Introduction}

Experiments with tagged photons, radiated from the initial state in
electron--proton and electron--positron collisions, can become particularly
attractive. The reason is that these radiative processes will permit to extract
information about the final states at continuously varying values of the
collision energy.
To investigate deep inelastic scattering the authors of Ref.~\cite{kps92}
suggested to use radiative events instead of running colliders at
reduced beam energies. The method takes advantage of a
photon detector (PD) placed in the very forward direction, as seen from the
incoming electron beam.
The effective beam energy, for each radiative event, is determined
by the energy of the hard photon is observed in PD. In fact, radiative
events were already used to measure the structure function $F_2$
down to $Q^2\geq 1.5$~GeV$^2$~\cite{h196}.
The specific theoretical work
concerns the evaluation of QED radiative
corrections~\cite{bkr97,tageps} to the radiative Born cross section.
With an accurate determination of the cross sections and of the possible
sources of background we believe that the use of radiative events may
become particularly useful to carry investigations at various present
and future machines.

The important role of the initial state radiation in the process
of electron--positron annihilation was underlined in a series of
papers by V.N.~Baier and V.A.~Khoze~\cite{BK65}, where the radiative
process was studied in detail in the Born approximation. In these
papers the mechanism of returning to a resonant region was
discovered. This mechanism consists in the preferable emission
of photons from the initial particles, which provides a
resonant kinematics of a subprocess.
A utilization of radiative events can become a common type of
investigations at various machines.

In this paper we derive explicit formulae for the spectrum
of tagged photons. The calculations are performed having
an accuracy of the per--mille order as an aim.
Formulae can be used at electron--positron
colliders to investigate, for instance, hadronic final states at intermediate
energies. A measurement of the total hadronic cross section at low energies
is essential for an high precision test of the Standard Model particularly
for a precise determination of the fine structure constant
$\alpha_{QED}(M_Z)$ and of the muon anomalous magnetic moment $(g-2)_{\mu}$.
The largest contribution to the errors for these
quantities comes from the large indetermination still present on the
measurement of the total hadronic cross section in electron--positron
annihilation at center--of--mass energies of a few GeVs. We will consider
here the radiatively corrected cross section for the electron--positron
annihilation process
\begin{equation}\label{1}
e^-(p_1)\ +\ e^+(p_2)\ \longrightarrow\ \gamma(k)\ +\ H(q), \qquad
k=(1-z)p_1 \, ,
\end{equation}
where $H$ is a generic hadronic state. The hard photon
hitting the photon detector has a momentum $k$ and an energy fraction $1-z$
with respect to the beam energy. In the following we assume that the photon
detector is placed along the electron beam direction, and has an opening
angle $2\theta_0 \ll 1$, such that $\varepsilon^2 \theta_0^2 \gg m^2$,
with $m$ the electron mass, and $\varepsilon$ the beam energy.
To evaluate the process with an accuracy of the per mille requires a careful
investigation of the radiative corrections. This paper is organized as follows.
In Section~2 we consider the cross section of the process~(\ref{1})
in the Born approximation. We give formulae suitable to study as
differential distributions in hadronic channels, as well as the total
(in terms of quantity $R$) and inclusive (in terms of hadron fragmentation
functions) hadronic cross sections. In Sec.~3 we calculate separate
contributions into radiatively corrected cross section of
process~(\ref{1}) within the next--to--leading accuracy. In Sec.~3.1 the
contribution due to virtual and soft photon emission is investigated.
In Sec.~3.2 the case, when additional hard photon hits a photon detector
is considered. In Sec.~3.3 the contribution due to hard photon emission,
which does not hit a photon detector is derived. In Sec.~4 we sum up all
the contributions and give the final result.
In Conclusions we summarize the results and give some numerical
illustrations.

\section{The Born approximation}

In order to obtain the Born approximation for the cross section of the
process~(\ref{1}),
when the PD is placed in front of electron (positron) beam,
we can use the quasireal electron method~\cite{bfk73}.
It gives
\begin{equation}\label{2}
\dd \sigma(k,p_1,p_2) = \dd W_{p_1}(k) \sigma_0(p_1-k,p_2),
\end{equation}
where $\dd W_{p_1}(k)$ is the probability to radiate photon
with energy fraction $1-z$ inside a narrow cone with the polar angle
not exceeding $ \theta_0 \ll 1$
around the incoming electron,
and $\dd \sigma_0$ is the differential cross section for the radiationless
process of electron--positron annihilation into hadrons at the
reduced electron beam energy. The form of both, $\dd W_{p_1}(k)$ and
$\sigma_0(p_1-k,p_2)$ is well known:
\begin{eqnarray}
\dd W_{p_1}(k) &=& \frac{\alpha}{2\pi}P_1(z,L_0)\dd z\, , \qquad
P_1(z,L_0) = \frac{1+z^2}{1-z}L_0 - \frac{2z}{1-z}\, ,\quad
L_0 = \ln\frac{\varepsilon^2\theta^2_0}{m^2}\, .
\end{eqnarray}
We need further the general form of the lowest order cross section
$\sigma_0$ for the process $e^+(z_1p_2) + e^-(zp_1) \to $ hadrons
boosted along the beam axis ($\vecc{p}_1$):
\begin{eqnarray} \label{3}
\sigma_0(z,z_1) &=& \frac{8\pi^2\alpha^2}{q^2|1-\Pi(q^2)|^2}
\int T(q) \dd\Gamma(q), \qquad
T(q) = \frac{L_{\rho\sigma}H_{\rho\sigma}}{(q^2)^2}, \\
L_{\rho\sigma} &=& \frac{q^2}{2}\widetilde{g}_{\rho\sigma}
+ 2z^2\widetilde p_{1\rho}\widetilde p_{1\sigma}, \qquad
\dd \Gamma(q) = (2\pi)^4\delta(q-\sum q_j)\prod\frac{\dd^3q_j}
{2\varepsilon_j(2\pi)^3} \, , \\ \nonumber
q &=& zp_1 + z_1p_2 \,\qquad q^2=sz_1z,\
\widetilde g_{\rho\sigma} = g_{\rho\sigma} -
\frac{q_{\rho}q_{\sigma}}{q^2}\, ,\qquad
\widetilde{p}_{1\rho} = p_{1\rho} - \frac{p_1q}{q^2}q_{\rho}\, ,
\end{eqnarray}
where $q$ is the full 4--momentum of final hadrons, $q_j$ is 4--momentum of
an individual hadron, $s = 2p_1p_2 = 4\varepsilon^2$ is the full
center--of--mass  energy squared, and $H_{\rho\sigma}$ is the hadronic
tensor. The vacuum polarization operator $ \Pi(q^2) $ of the virtual
photon with momentum $q$ is a known function~\cite{bw96} and will not be
specified here.

The tensors $H_{\rho\sigma}$ and $L_{\rho\sigma}$ obey the current
conservation conditions once saturated with the 4--vector $q$.
The differential cross section with respect to the tagged photon energy
fraction $z$ can be obtained by performing the integration on the hadrons
phase space. It takes the form
\begin{equation}\label{4}
\frac{\dd \sigma}{\dd z} = \frac{\alpha}{2\pi}P_1(z,L_0)\;\sigma_0(z,1).
\end{equation}

Each hadronic state is described by its own hadronic tensor. The
cross section in Eqs.~(3) and (4) is suitable for different uses and, as
mentioned above, it can be used to check different theoretical predictions.

The sum of the contributions of all hadronic
channels by means of the relation
\begin{equation}\label{6}
\sum_{h}\int H_{\rho\sigma}\dd\Gamma = f_h(q^2)
\widetilde g_{\rho\sigma}\, ,
\end{equation}
can be expressed
in terms of the ratio of the total cross section for annihilation
into hadrons and muons $R = \sigma_h/\sigma_{\mu}$.
For the $\mu^+\mu^-$ final state we get
\begin{eqnarray*}
f_{\mu} = \frac{q^2}{6\pi}K(q^2), \qquad
K(q^2) = \biggl(1+\frac{2m^2_{\mu}}{q^2}\biggr)
\sqrt{1-\frac{4m^2_{\mu}}{q^2}},
\end{eqnarray*}
and so,
\begin{equation}\label{7}
  f_h(q^2) = \frac{q^2R(q^2)}{6\pi}K(q^2).
\end{equation}
Substituting this expression into the right hand side
of Eqs.~(3,4) results in the replacement $\sigma_0(z,z_1)
=R(q^2)4\pi\alpha^2 K(q^2)/(3q^2)$.

In experiments of semi--inclusive type one fixes an hadron with 3-momentum
$\vecc{q}_1$ energy $\epsilon_1$ and mass $ M $ in every
event and sum over all the rest. In this case instead of Eq.~(\ref{6})
we will have (similarly to the Deep Inelastic Scattering (DIS)
case~\cite{tageps}):
\begin{eqnarray}\label{9}
&& \sum_{h'}\int H_{\rho\sigma}\dd \Gamma =
H^{(1)}_{\rho\sigma}\frac{\dd^3 q_1}{2\varepsilon_1(2\pi)^3} \, ,
\\ \nonumber
&& H^{(1)}_{\rho\sigma} = F_1(\eta,q^2)\widetilde g_{\rho\sigma} -
\frac{4}{q^2} F_2(\eta,q^2)\widetilde q_{1\rho}\widetilde q_{1\rho}, \qquad
 \eta = \frac{q^2}{2qq_1}>1 \, ,
\end{eqnarray}
where we have introduced two dimentionless functions $F_1(\eta,q^2)$ and
$F_2(\eta,q^2)$ in a way similar to the DIS case.

By introducing the dimentionless variable $\lambda = 2qq_1/(2zp_1q_1)$,
we can write the corresponding cross section for radiative events in
$e^+e^-$ annihilation in the same form as in the case of
deep inelastic scattering with a tagged photon \cite{tageps}:
\begin{eqnarray}\label{10}
\frac{\dd \sigma}{\dd z} &=& \frac{\alpha^2(q^2)}{2\pi}\,\frac{\alpha}{2\pi}
P_1(z,L_0)\Sigma(\eta,\lambda,q^2)\frac{1}{(q^2)^2}\,
\frac{\dd^3 q_1}{\varepsilon_1}\, ,
\\ \nonumber
\Sigma(\eta,\lambda,q^2) &=&
F_1(\eta,q^2) + \frac{2F_2(\eta,q^2)}{\eta^2\lambda^2}
\biggl(\lambda - 1 - \frac{M^2}{q^2}\eta^2\lambda^2\biggr).
\end{eqnarray}

\section{Radiative corrections}

For the radiative corrections (RC) to the cross section~(5)
we will restrict ourselves only to terms containing second and first
powers of large logarithms $L$, and omit terms which don't contain
them i.e. we will keep leading and next--to--leading logarithmic
contributions. We will consider in subsection~3.1 the contribution
from one--loop virtual photon as well as from the emission of soft real ones. In~3.2
we will discuss the double hard photon emission process.

\subsection{Corrections due to the virtual and real soft photons}

The interference of Born and one--loop contributions to the amplitude of
the initial state radiation in annihilation of $e^+e^-$ into hadrons
can be obtained from the analogous quantity of hard
photon emission in electron--proton scattering~\cite{tageps}.
We do that by using the crossing transformation.
For the contribution coming from the emission of real soft
photons a straightforward calculation gives:
\begin{eqnarray}
\frac{\dd\sigma^{S}}{\dd\sigma_0}=\frac{\alpha}{\pi}\biggl[2(L_s-1)
\ln\frac{m \Delta \varepsilon}{\lambda \varepsilon}+\frac{1}{2}L_s^2-
\frac{\pi^2}{3}\biggr], \quad
L_s=\ln\frac{s}{m^2} = L_0 + L_{\theta}, \quad
L_{\theta} = \ln\frac{4}{\theta^2}\, .
\end{eqnarray}
where $\lambda$ is the {\it photon mass}, $\Delta\varepsilon$ is the energy
carried by the soft photon. The sum of the two contributions
is free from infrared singularities. It reads
\begin{equation}
\dd \sigma^{V+S}=\frac{8\pi^2\alpha^2}{s|1-\Pi(q^2)|^2}\frac{\alpha}{2\pi}
[\rho B_{\rho\sigma}(q)+A_{\rho\sigma}(q)]\frac{H_{\rho\sigma}(q)
\dd \Gamma(q)}{(q^2)^2}\frac{\alpha}{4\pi^2}\frac{\dd^3 k}{\omega},
\end{equation}
where
\begin{equation}
\rho=4(L_s-1)\ln \Delta+3L_q-\frac{\pi^2}{3}-\frac{9}{2},\quad
L_q=L_s+\ln z,\quad
\Delta=\frac{\Delta\varepsilon}{\varepsilon}\ll 1,
\end{equation}
where $k$ and $\omega$ are the 3-momentum and the energy of the hard photon
respectively. The tensors
$A_{\rho\sigma}$ and $B_{\rho\sigma}$ have a rather involved form. The first
can be obtained from the corresponding expressions of Ref.~\cite{kmf87}. The
tensor $B_{\rho\sigma}$ coincides with the one of the Born approximation.
In the kinematical region where the hard photon is emitted close to the
initial electron direction of motion one has
\begin{equation}
B_{\rho\sigma}=\frac{2}{z}\biggl( \frac{1+z^2}{y_1(1-z)}
- \frac{2m^2z}{y_1^2} \biggr)L_{\rho\sigma}(q), \qquad
A_{\rho\sigma}=\frac{2}{q^2}A_gL_{\rho\sigma}(q), \quad
q=zp_1+p_2,
\end{equation}
where tensor $L_{\rho\sigma}$ is given in Eq.~(4),
$y_1=2kp_1$, and quantity $A_g$ reads
\begin{eqnarray}
A_g&=&\frac{4 z s m^2}{y_1^2}L_s \ln z +\frac{s}{y_1}\biggl[
\frac{1+z^2}{1-z}(-2L_s \ln z
-\ln^2 z+2\Li(1-z) \nonumber\\
&+&2\ln\frac{y_1}{m^2}\ln z)+\frac{1+2z-z^2}{2(1-z)}\biggr],\qquad
\Li(x)=-\int_0^1 \dd t \frac{\ln(1-tx)}{t}.
\end{eqnarray}
Further integration over the hard photon phase space can be performed
within the logarithmic accuracy by using the integrals
\begin{equation}
\int\frac{\dd^3k}{2\pi k_0}\biggl[\frac{1}{y_1},\frac{m^2}{y_1^2},
\frac{\ln(y_1/m^2)}{y_1}\biggr]=\biggl[\frac{1}{2}L_0,\frac{1}{2(1-z)},
\frac{1}{4}L_0^2+
\frac{1}{2}L_0\ln(1-z)\biggr]\dd z. \nonumber
\end{equation}
The final expression for the Born cross section corrected for the
emission of soft and virtual photons has the form
\begin{eqnarray}
\frac{\dd \sigma^{B+V+S}}{\dd z}&=&\sigma_0(z,1)\biggl[
\frac{\alpha}{2\pi}P_1(z,L_0)+
\biggl(\frac{\alpha}{2\pi}\biggr)^2(\rho P_1(z,L_0)+N)\biggr], \nonumber\\
N&=&-\frac{1+z^2}{1-z}\biggl[(L_0+\ln z)\ln z - \frac{\pi^2}{3}
+ 2\Li(z)\biggr]L_0 - 2P_1(z,L_0)\ln\frac{\theta_0^2}{4}\nonumber \\
&+& \frac{1+2z-z^2}{2(1-z)}L_0+\frac{4z}{1-z}L_0\ln z.
\end{eqnarray}

\subsection{Two hard photons are tagged by the detector}

If an additional hard photon emitted by the initial--state electron
hits the PD, we cannot use the quasireal electron method and have
to calculate the corresponding contribution starting from
Feynman diagrams.

We can use double hard photon spectra as given in Ref.~\cite{xji}
for annihilation diagrams only and write the
cross section under consideration as follows
\begin{eqnarray}
\frac{\dd \sigma^H_{c1}}{\dd z} &=& \sigma_0(z,1)
\biggl(\frac{\alpha}{2\pi}\biggr)^2 L_0\int\limits_{\Delta}^{1-z-\Delta}
\frac{\dd x}{\xi}
\biggl[\frac{\gamma\tau}{2}L_0 + (z^2 + (1-x)^4)\ln\frac{(1-x)^2(1-z-x)}{zx}
\nonumber \\ \label{25}
&+& zx(1-z-x) - x^2(1-x-z)^2 -2\tau(1-x)\biggr], \\ \nonumber
\xi &=& x(1-x)^2(1-z-x), \quad \gamma = 1+(1-x)^2, \quad
\tau = z^2 + (1-x)^2.
\end{eqnarray}
Here the variable $x$ under the integral sign is the energy fraction of
one hard photon. The quantity $1-z-x$ is the energy fraction of the second
hard photon provided that their total energy fraction equals to $1-z$.
We write the index $c1$ in the left hand side of Eq.~(\ref{25})
to emphasize that this contribution
arises from the collinear kinematics, when the additional hard
photon is emitted along the initial electron with 4--momentum $p_1$.

The integration in the right hand side of Eq.~(\ref{25}) leads to the result
\begin{eqnarray}
\frac{\dd \sigma^{H}_{c1}}{\dd z} &=& \sigma_0(z,1)
\biggl(\frac{\alpha}{2\pi}\biggr)^2\frac{L_0}{2}\biggl\{\biggl[
P^{(2)}_{\Theta}(z) + 2\frac{1+z^2}{1-z}
\biggl(\ln z - \frac{3}{2} - 2\ln\Delta\biggr)\biggr]L_0
\nonumber \\ \label{26}
&+& 6(1-z) + \frac{3+z^2}{1-z}\ln^2z
- \frac{4(1+z)^2}{1-z}\ln\frac{1-z}{\Delta} \biggr\},
\end{eqnarray}
where the quantity $P^{(2)}_{\Theta}(z)$ represents the so called
$\Theta$--term of the second--order electron structure function:
\begin{eqnarray}
P^{(2)}_{\Theta}(z) = 2\frac{1+z^2}{1-z}\biggl(\ln\frac{(1-z)^2}{z}
+ \frac{3}{2}\biggr) + (1+z)\ln z - 2(1-z).
\end{eqnarray}

\subsection{Additional hard photon is emitted outside PD}

If an additional hard photon, emitted from the initial state, does not hit
the PD situated in the direction of motion of the initial electron
we distinguish the case when it is emitted in the direction close,
within a small cone with angle $ \theta'\ll 1 $, to the direction
of the initial positron. In this case we obtain:
\begin{equation}
\frac{\dd\sigma^H_{c2}}{\dd z}=\frac{\alpha}{2\pi}P_1(z,L_0)
\int_{\Delta}^{1-
\delta/z}\frac{\alpha}{2\pi}P_1(1-x,L')\sigma_0(z,1-x)\dd x,
\end{equation}
where $ L'=L_s+\ln(\theta'^2/4)$, $\delta=M^2/s$,
and $ M^2$ is the minimal hadron mass squared. We suppose that
$z \sim 1$.

We have introduced the additional auxiliary parameter $\theta' \ll 1$
which, together with $\theta_0$, separates collinear and semi--collinear
kinematics of the second hard photon. Contrary to $\theta_0$,
which is supposed to determine the PD acceptance, $\theta'$
will disappear in the sum of the collinear and semi--collinear contributions
of the second photon. This last kinematical region gives
\begin{eqnarray}
\frac{\dd\sigma^H_{sc}}{\dd z}&=&(\frac{\alpha}{2\pi})^2 P_1(z,L_0)
\int \frac{\dd^3 k_1}{2\pi\omega_1^3}\frac{16\pi^2\alpha^2}{(1-c^2)z^2}
T(c,z,x),\nonumber\\
T(c,z,x)&=&\int \frac{H_{\rho\sigma}(q_2)\dd \Gamma(q_2)}
{s(q_2^2)^2|1-\Pi(q_2^2)|^2}\biggl[\frac{s}{2}((z-x_2)^2+z^2(1-x_1)^2
)g_{\rho\sigma} \nonumber\\
&+&2(z(1-x_1)-x_2)(z^2p_{1\rho}p_{1\sigma}+
p_{2\rho}p_{2\sigma})\biggr],\qquad
x_1=\frac{x}{2}(1-c),\quad x_2=\frac{x}{2}(1+c),\nonumber\\
q_2&=&zp_1+p_2-k_1, \qquad c=\cos \widehat{\vecc{k}_1\vecc{p}_1}.
\end{eqnarray}
The phase volume of the second photon is parametrized as:
\begin{equation}
\int \frac{\dd^3 k_1}{2\pi\omega^3}=\int\limits_{\Delta}^{\hat x}
\frac{\dd x}{x}
\int\limits_{0}^{2\pi}\frac{\dd \phi}{2\pi}
\int\limits_{-1+\theta'^2/2}^{1-\theta_0^2/2}
\dd c,  \qquad
\hat x=\frac{2(z-\delta)}{1+z+c(1-z)}.
\end{equation}
Explicitly extracting the angular singularities we represent this
expression as
\begin{eqnarray}
\frac{\dd\sigma^H_{sc}}{\dd z}&=&\biggl(\frac{\alpha}{2\pi}\biggr)^2
P_1(z,L_0)\biggl[
\Sigma_{sc}(z)+
\ln\frac{4}{\theta_0^2}
\int\limits_{\Delta}^{z-\delta}\frac{\dd x}{x}
\frac{z^2+(z-x)^2}{z^2}\sigma_0(z-x,1) \nonumber\\
&+&\ln\frac{4}{\theta'^2} \int\limits_{\Delta}^{1-\delta/z}
\frac{\dd x}{x}(1+(1-x)^2)
\sigma_0(z,1-x)\biggr], \\ \nonumber
\Sigma_{sc}&=&\frac{8\pi^2\alpha^2}{z^2}\int\limits_{-1}^{1}\dd c
\int\limits_{\Delta}^{\hat x}
\frac{\dd x}{x}
\biggl[
\frac{T(c,z,x)-T(1,z,x)}{1-c}+\frac{T(c,z,x)-T(-1,z,x)}{1+c}\biggr].
\end{eqnarray}

\section{Complete QED correction and leading logarithmic approximation}

The final result in the order ${\cal O}(\alpha)$ for radiative
corrections to radiative events can be written as follows:
\begin{eqnarray} \label{final}
\frac{\dd\sigma}{\dd z} &=& \frac{\alpha}{2\pi}P_1(z,L_0)\sigma_0(z,1)
( 1 + r ) \\ \nonumber
&=& \frac{\alpha}{2\pi}P_1(z,L_0)\sigma_0(z,1)
+ \biggl(\frac{\alpha}{2\pi}\biggr)^2
\Biggl\{L_0\biggl(\frac{1}{2}L_0P^{(2)}(z) + G\biggr)\sigma_0(z,1)
\\ \nonumber
&+& P_1(z,L_0)\biggl[\int
\limits_{0}^{1-\delta/z}C_1(x)\sigma_0(z,1-x) \dd x
+ L_{\theta}\int\limits_{0}^{z-\delta}C_2(z,x)
\sigma_0(z-x,1) \dd x + \Sigma_{sc}\biggr]\Biggr\},
\end{eqnarray}
where the last term is defined by Eq.~(23) and
\begin{eqnarray} \nonumber
C_1(x) &=& P_1(1-x,L_s)\Theta(x-\Delta)
+ (L_s-1)(2\ln\Delta + \frac{3}{2})\delta(x), \\ \label{36}
C_2(z,x) &=& \frac{z^2+(z-x)^2}{z^2x}\Theta(x-\Delta)
+ ( 2\ln\Delta + \frac{3}{2} - 2\ln z)\delta(x),
\\ \nonumber
G(z) &=& \frac{1+z^2}{1-z}(3\ln z - 2 \Li(z)) + \frac{1}{2}(1+z)\ln^2z
- \frac{2(1+z)^2}{1-z}\ln(1-z) \\ \nonumber
&+& \frac{1-16z-z^2}{2(1-z)} + \frac{4z\ln z}{1-z} \, .
\end{eqnarray}

In order to include the higher order leading corrections to the tagged
photon differential cross--section and show the agreement of our
calculation with the well known Drell--Yan representation for the total
hadronic cross--section at electron--positron annihilation~\cite{kf85}
\begin{equation} \label{kf}
\sigma(s) = \int\limits_{\delta}^1\dd x_1\int\limits_{\delta/x_1}^1\dd x_2\;
D(x_1,\alpha_{\mathrm{eff}})D(x_2,\alpha_{\mathrm{eff}}) \sigma(x_1x_2s),
\end{equation}
where the electron structure functions include both nonsinglet and singlet
parts
\begin{equation} \label{sf} 
D(x_1,\alpha_{\mathrm{eff}}) =  D^{NS}(x,\alpha_{\mathrm{eff}})
+ D^S(x_1,\alpha_{\mathrm{eff}}),
\end{equation}
it is convenient to introduce the quantity
\begin{equation}\label{LLA}
\Sigma =
D(z,\bar\alpha_{\mathrm{eff}})
\int\limits_{\delta/z}^1\dd x_1\int\limits_{\delta/zx_1}^1\dd x_2\;
D(x_1,\widetilde\alpha_{\mathrm{eff}})D(x_2,\hat\alpha_{\mathrm{eff}})
\sigma_0(zx_1,x_2).
\end{equation}
Note that the shifted cross--section in Eq.~(\ref{kf}) has just the same
meaning as in Eq.~(\ref{3}): $\sigma(x_1x_2s)=\sigma_0(x_1,x_2)$.

The structure functions~\cite{jsw,nt}, which enter into the right side
of Eq.~(\ref{sf}), are
\begin{equation}\label{30}
D^{NS}(x,\alpha_{\mathrm{eff}}) = \delta(1-x)
+ \sum_{n=1}^{\infty}\frac{1}{n!}\Biggl(\frac{\alpha_{\mathrm{eff}}}{2\pi}
\Biggr)^n P_1^{\otimes n}(x),
\end{equation}
\begin{equation}\label{31}
D^S(x,\alpha_{\mathrm{eff}}) =
\frac{1}{2!}\Biggl(\frac{\alpha_{\mathrm{eff}}}{2\pi}\Biggr)^2 R(x)
+ \frac{1}{3!} \Biggl(\frac{\alpha_{\mathrm{eff}}}{2\pi}\Biggr)^3
\biggl[2P_1\otimes R(x) - \frac{2}{3}R(x)\biggr],
\end{equation}
where
\begin{eqnarray*}
P_1(x) &=& \lim_{\Delta\to 0}\biggl\{\frac{1+x^2}{1-x}\Theta(1-\Delta-x)
+ \biggl(\frac{3}{2}+2\ln\Delta\biggr)\delta(1-x)\biggr\}, \\
R(x) &=& 2(1+x)\ln x + \frac{1-x}{3x}(4+7x+4x^2), \\
P_1^{\otimes n} &=& \underbrace{P_1(x)\otimes\cdots\otimes P_1(x)}_n,
\qquad P_1(x)\otimes P_1(x) =
\int\limits_x^1P_1(t)P_1\biggl(\frac{x}{t}\biggr)\frac{\dd t}{t},
\end{eqnarray*}
and
the effective electromagnetic couplings in the right side
of Eq.~(\ref{LLA}) are
\begin{equation}\label{32}
\bar\alpha_{\mathrm{eff}} = -3\pi\ln\biggl(1-\frac{\alpha}{3\pi}L_0\biggr),
\quad \widetilde\alpha_{\mathrm{eff}} =
-3\pi\ln\biggl(\frac{1-\frac{\alpha}{3\pi}L_s}{1-\frac{\alpha}{3\pi}
L_0}\biggr),
\quad \hat\alpha_{\mathrm{eff}} =
-3\pi\ln\biggl(1-\frac{\alpha}{3\pi}L_s\biggr).
\end{equation}

At fixed values of $z$ $(z<1)$ the quantity $\Sigma$ defines the leading
logarithmic contributions into differential cross--section for
the events with tagged particles. That corresponds to only
$\Theta$--terms in the
expansion of the structure function
$D(z,\bar\alpha_{\mathrm{eff}})$ before the
integral sign in Eq.~(\ref{LLA}). If we consider photonic corrections (as in
previous Sections) it needs to restrict ourselves with the nonsinglet part
of the electron structure functions and with the first order terms in the
expansion of all effective couplings, namely:
\begin{equation}\label{33}
\bar\alpha_{\mathrm{eff}} \rightarrow \alpha L_0, \quad
\widetilde\alpha_{\mathrm{eff}} \rightarrow \alpha L_{\theta}, \quad
\hat\alpha_{\mathrm{eff}} \rightarrow \alpha L_s.
\end{equation}
It is easy to see that in this case the leading
contribution into differential cross--section~(\ref{final})
can be obtained as an
expansion of the quantity $\Sigma(z<1)$ by the powers of $\alpha$, keeping
the terms of the order $\alpha^2$ in the production of $D$--functions.

If we want to include the contribution due to $e^+e^-$--pair (real and
virtual) production it is required~\cite{ct} to use both nonsinglet
and singlet
structure functions and effective couplings defined by Eq.~(\ref{32}).
Note that the insertion into consideration of higher order corrections
rises additional questions about concrete experimental conditions
concerning registration of events with $e^+e^-$--pairs.

The total hadronic cross--section in $e^+e^-$ annihilation can be obtained
by integration of quantity $\Sigma$ over $z$
\begin{equation}\label{tots}
\sigma(s) = \int\limits_{\delta}^1\dd z\; D(z,\bar\alpha_{\mathrm{eff}})
\int\limits_{\delta/z}^1\dd x_1\; \int\limits_{\delta/zx_1}^1\dd x_2\;
D(x_1,\widetilde\alpha_{\mathrm{eff}})D(x_2,\hat\alpha_{\mathrm{eff}})
\sigma(zx_1x_2s).
\end{equation}
We can integrate the expression in the right side of Eq.~(\ref{tots})
over the variable $z$ provided the quantity $zx_1 =y$ is fixed
\begin{eqnarray}\label{35}
&& \int\limits_{\delta}^1\dd z\; D(z,\bar\alpha_{\mathrm{eff}})
\int\limits_{\delta/z}^1 \dd x_1\; D(x_1,\widetilde\alpha_{\mathrm{eff}})
= \int\limits_{\delta}^1\dd z\int\limits_y^1\dd y\;
D(z,\bar\alpha_{\mathrm{eff}})
D\biggl(\frac{y}{z}, \widetilde\alpha_{\mathrm{eff}}\biggr) \\ \nonumber
&& \qquad =
\int\limits_{\delta}^1\dd y\; D(y,\bar\alpha_{\mathrm{eff}}
+ \widetilde\alpha_{\mathrm{eff}}), \quad
\bar\alpha_{\mathrm{eff}} + \widetilde\alpha_{\mathrm{eff}}
= \hat\alpha_{\mathrm{eff}}.
\end{eqnarray}
Using this result and definition of $\hat\alpha_{\mathrm{eff}}$ we indicate
the equivalence of the Drell--Yan form of the total cross--section as given
by Eq.~(\ref{kf}) and the representation of the cross--section by
Eq.~(\ref{tots}).

Let us show now that $D$--functions in expression for the quantity
$\Sigma$ have effective couplings as given by Eq.~(\ref{32}).
By definition the
nonsinglet electron structure function satisfies the equation~\cite{LAP}
\begin{equation} \label{lap}
D(x,s,s_0) = \delta(1-x) +
\frac{1}{2\pi}\int\limits_{s_0}^s\frac{ds_1}{s_1}\alpha(s_1)\int\limits_
x^1\frac{dz}{z}D(z)D\biggl(\frac{x}{z},\frac{s_1}{s_0}\biggr),
\end{equation}
where $\alpha(s_1)$ is the electromagnetic running coupling
\begin{eqnarray*}
\alpha(s_1) =
\alpha\biggl(1-\frac{\alpha}{3\pi}\ln\frac{s_1}{m^2}\biggr)^{-1},
\end{eqnarray*}
and $s_0(s)$ is the minimal (maximal) virtuality of the particle, which
radiate photons and $e^+e^-$--pairs.

The structure function $D(z,\bar\alpha_{\mathrm{eff}})$ describes the photon
emission and pair production inside narrow cone along the electron beam
direction. In this kinematics $s_0 = m^2$, $s= \varepsilon^2\theta_0^2$.
The corresponding iterative solution of the Eq.~(\ref{lap})
has the form~(\ref{30}) with
$\alpha_{\mathrm{eff}} = \bar\alpha_{\mathrm{eff}}$. The structure function
$D(x_1,\widetilde\alpha_{\mathrm{eff}})$ describes the events,
when emitted (by the electron) particles escape this narrow cone.
In this case $s_0 = \varepsilon^2\theta_0^2$, $\ s= 4\varepsilon^2$. The
corresponding solution of Eq.~(\ref{lap}) gives the structure function with
$\alpha_{\mathrm{eff}} = \widetilde\alpha_{\mathrm{eff}}$.
At last, the structure function $D(x_2,\hat\alpha_{\mathrm{eff}})$
is responsible for the radiation off the positron into the whole phase space.
In this case $s_0=m^2$, $\ s=4\varepsilon^2$.
Therefore we obtain $D$--function with $\alpha_{\mathrm{eff}} =
\hat\alpha_{\mathrm{eff}}$. The analogous consideration can be performed for
the singlet part of structure functions.

When writing the representation~(\ref{tots}) for the total cross--section we, in
fact, divide the phase space of the particles emitted by the electron on
the regions inside and outside the narrow cone along electron beam
direction. Therefore we can use this representation to investigate the
events with tagged particles in both this regions. As we saw before the
differential cross--section for events with tagged particles inside the
narrow cone is defined by the quantity $\Sigma(z<1)$. In order to obtain
the corresponding differential cross--section for events with tagged
particles outside this narrow cone we have to change the places of
$\bar\alpha_{\mathrm{eff}}$ and $\widetilde\alpha_{\mathrm{eff}}$
in expression for $\Sigma(z,1)$.
This follows from the symmetry of representation~(\ref{tots})
relative such change.

\section{Conclusion}

In sum the formulae~(\ref{35},\ref{LLA}) are the main results of our paper.
In Fig.~1 we show the cross section $\dd\sigma/\dd z$ as a function of $z$.
The beam energy is chosen to be $E_{\mathrm{beam}}=0.5$~GeV.
The region of $z$ values is limited by the pion production threshold
at the left, and by the threshold of photon detection (we choose 50~MeV)
at the right. The peak in middle corresponds to the
large contribution of the $\rho$-meson. Values of $R$, used for numerical
estimations, were taken from~\cite{Dol}.
The values of corrections $r$
(see Eq.~(\ref{35})) in percent are shown in Fig.~2.

\begin{figure}[t]
\begin{center}
\mbox{\epsfig{file=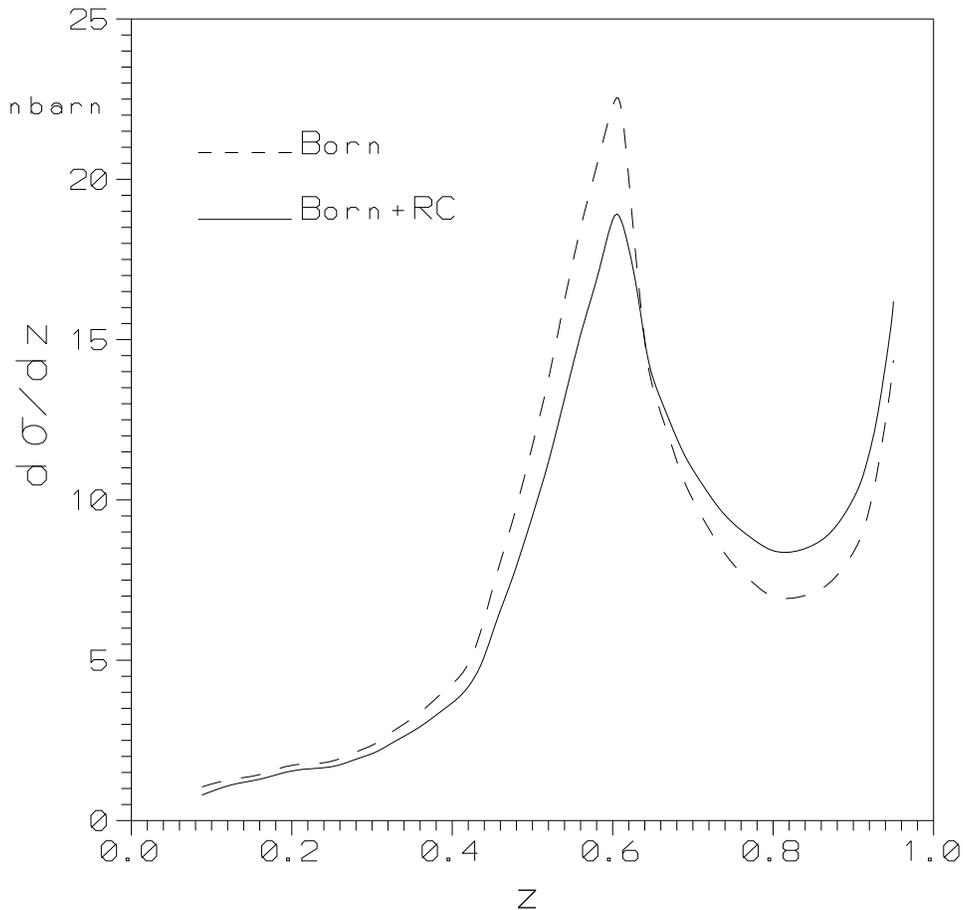,height=12cm,angle=0}}
\end{center}
\caption{ The cross section of $e^+e^-\to$ hadrons\ with
tagged photon.}
\end{figure}

\begin{figure}[t]
\begin{center}
\mbox{\epsfig{file=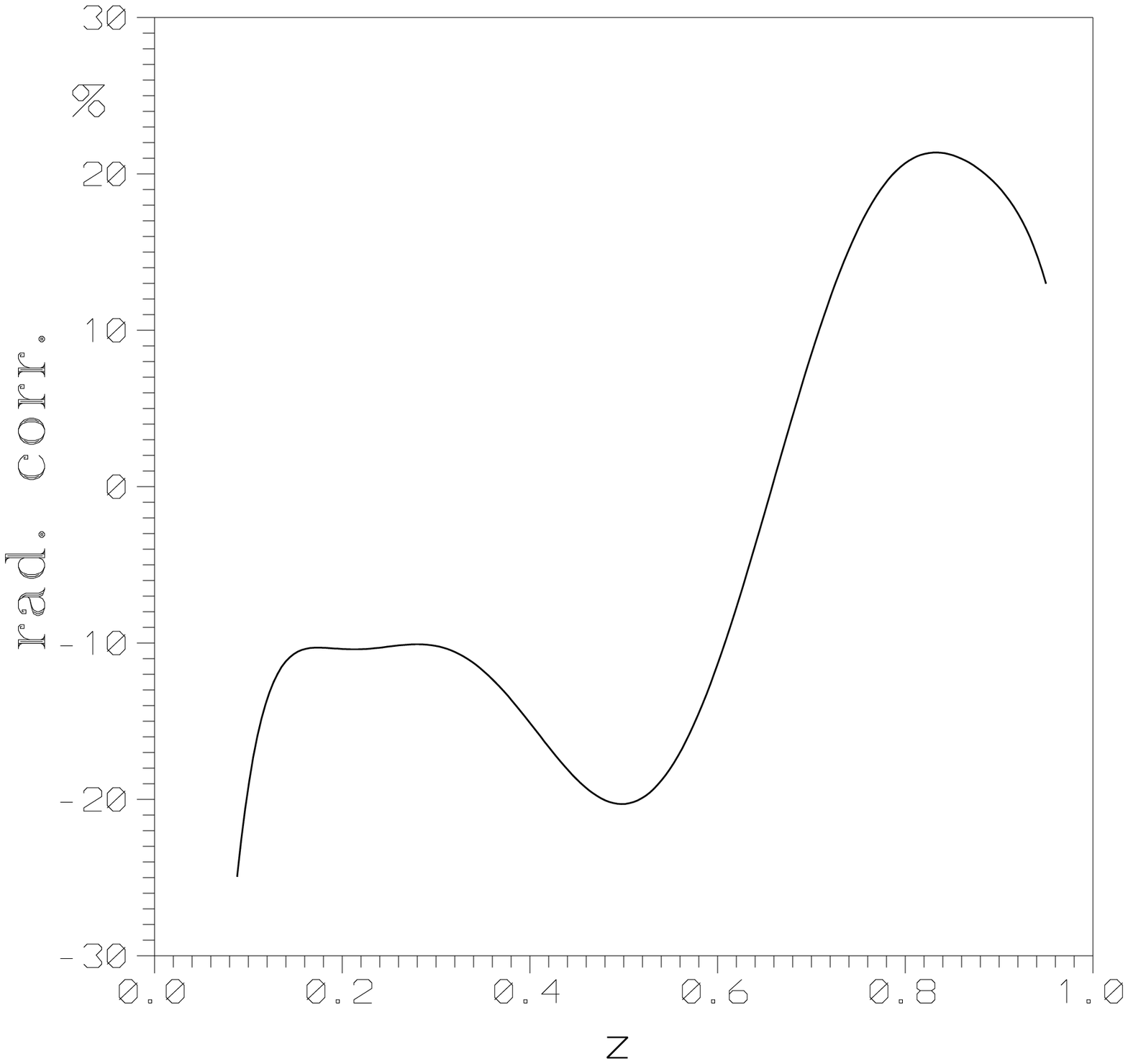,height=12cm,angle=0}}
\end{center}
\caption{ The radiative corrections to $e^+e^-\to$ hadrons\ with
tagged photon.}
\end{figure}


So, we calculated the cross--section of $e^+e^-$ annihilation
with detection of a hard photon at small angles in respect to
the electron beam. The general structure of a measured cross--section,
from which one should extract the annihilation crossection $\sigma_0$,
looks as follows:
\begin{eqnarray}
\sigma = \sigma_0\biggl[ a_1\frac{\alpha}{\pi}L + b_1\frac{\alpha}{\pi}
+ a_2\biggl(\frac{\alpha}{\pi}\biggr)^2L^2
+ b_2\biggl(\frac{\alpha}{\pi}\biggr)^2L
+ c_2\biggl(\frac{\alpha}{\pi}\biggr)^2 \biggr] + {\cal O}(\alpha^3),
\end{eqnarray}
where $L$ denotes some large logarithm.
We calculated the terms $a_1$, $b_1$, $a_2$, $b_2$ and some contributions
to $c_2$. The generalized formula~(\ref{LLA}) allows to involve
the leading terms of the order ${\cal O}(\alpha^3L^3)$.
In this way our formulae provide high theoretical precision.

Similar formulae can be obtained for an experimental set--up
by tagging a definite hadron.
By using $e^+e^-$--machines such as BEPS, DA$\Phi$NE~\cite{s}, VEPP,
CLEO, SLAC--B/factory and others with luminosities of order
$10^{33}$~cm$^2 s^{-1}$, one may be in principle able to scan,
by measuring the initial state radiation spectrum,
the whole energy region of hadron production with an effective
luminosity of the order of $10^{31}$~cm$^2 s^{-1}$. We hope further study
will follow on these issues both from the experiments and theory.

\subsection*{Acknowledgments}

We are grateful to S.~Serednyakov and Z.~Silagadze for a critical reading
of the manuscript and for useful comments.
A.A., E.K. and N.P.M. wish to thank the INFN and the Dipartimento di Fisica
of the Universit\`a for the hospitality. This work is supported by the INTAS
grant 93--1867 ext. One of us (N.P.M.) thanks the Ukrainian
DFFD for the grant N24/379 and the Ministry of Science and Technology of
Ukraine for support. We wish also to thank B.~Shaikhatdenov for help.

\end{document}